\documentclass[12pt,preprint]{aastex}

\slugcomment{To appear in ApJ.}

\shorttitle{Correlation of energetic cosmic rays and gamma-ray sources}
\shortauthors{Nemmen, Bonatto \& Storchi-Bergmann}

\begin{document}

\title{A correlation between the highest energy cosmic rays and nearby active galactic nuclei detected by Fermi}

\author{Rodrigo S. Nemmen\altaffilmark{1}, Charles Bonatto and Thaisa Storchi-Bergmann}
\affil{Instituto de F\'isica, Universidade Federal do Rio Grande do Sul, Campus do Vale, Porto Alegre, RS, Brazil}

\altaffiltext{1}{E-mail: rodrigo.nemmen@ufrgs.br.}

\begin{abstract}
We analyze the correlation of the positions of $\gamma$-ray sources in the \textit{Fermi} Large Area Telescope First Source Catalog (1FGL) and the First LAT Active Galactic Nuclei (AGN) Catalog (1LAC) with the arrival directions of ultra-high-energy cosmic rays (UHECRs) observed with the \textit{Pierre Auger Observatory}, in order to investigate the origin of UHECRs.
We find that Galactic sources and blazars identified in the 1FGL are not significantly correlated with UHECRs, while the 1LAC sources display a mild correlation ($2.6\sigma$ level) on a $\approx 2.4^\circ$ angular scale.
When selecting only the 1LAC AGNs closer than 200 Mpc, we find a strong association ($5.4\sigma$) between their positions and the directions of UHECRs on a $\approx 17^\circ$ angular scale; the probability of the observed configuration being due to an isotropic flux of cosmic rays is $5 \times 10^{-8}$. There is also a $5\sigma$ correlation with nearby 1LAC sources on a $6.5^\circ$ scale.
We identify 7 ``$\gamma$-ray loud'' AGNs which are associated with UHECRs within $\approx 17^\circ$ and are likely candidates for the production sites of UHECRs: Centaurus A, NGC 4945, ESO 323-G77, 4C+04.77, NGC 1218, RX J0008.0+1450 and NGC 253.  
We interpret these results as providing additional support to the hypothesis of the origin of UHECRs in nearby extragalactic objects. As the angular scales of the correlations are large, we discuss the possibility that intervening magnetic fields might be considerably deflecting the trajectories of the particles on their way to Earth. 
\end{abstract}


\section{Introduction}	\label{sec:intro}

The nature of the ultra-high-energy cosmic rays (UHECRs) with energies above $\sim 10^{20}$ eV remains enigmatic since they were first observed, more than half a century ago \citep{linsley63}. If such cosmic rays are composed predominantly of protons or nuclei, the Greisen, Zatsepin and Kuzmin (GZK) effect \citep{g,zk} restricts their possible sources to the nearby universe, closer than about $\sim 100-200$ Mpc from Earth (the ``GZK horizon'', \citealt{harari06}). Under the assumption that the UHECRs sources are relatively nearby and not uniformly distributed on the sky, then an anisotropic  distribution of arrival directions is expected, as long as the deflections caused by the intervening magnetic fields are sufficiently small.

With the unprecedented capabilities of the \textit{Pierre Auger Observatory} (PAO; \citealt{pafirst}), it is now possible to map the arrival directions of UHECRs with a $\approx 1^\circ$ precision \citep{pa}. This makes the cross-correlation analysis of the arrival directions with catalogs of astronomical objects a powerful tool for tracking the UHECR production sites \citep{pa, aublin09, paupdate}. 
For instance, the anisotropy in the arrival directions of UHECRs was recently demonstrated with a significance level above 99\%, through the correlation between the directions of such particles and the positions of nearby active galactic nuclei (AGNs) from the V\'eron-Cetty and V\'eron catalog \citep{pa, palong}. This indicates that the highest-energy particles have an extragalactic origin within the GZK horizon, with either AGNs or other objects with a similar spatial distribution being the likely astrophysical sources of UHECRs. Note that the High Resolution Fly's Eye (HiRes) experiment does not confirm the correlations suggested by the PAO data \citep{hires08,hires10}.

Since the pioneering work of \citet{pa}, several authors have been studying the correlation of the PAO events' directions with different classes of astrophysical objects. There is circumstantial evidence that at least some UHECRs appear to be associated with relatively few nearby radio galaxies \citep{nagar08,moska09}, with a puzzling paucity of UHECRs in the direction of the Virgo cluster \citep{zaw09}. There is also evidence that spiral galaxies seem to host the producers of ultra-high energy cosmic rays \citep{guise08}. \citet{moska09} noted the importance of taking into account the AGN morphology in correlation studies. Moskalenko et al. also pointed out the possibility of UHECRs being correlated with larger deflection angles and/or more distant sources than considered by \citet{pa}. An understanding of which particular class of extragalactic sources is associated with the production sites of UHECRs is clearly missing.

Astrophysical $\gamma$-ray sources are appealing candidate accelerators of UHECRs, since by their very nature they are nonthermal objects where large energy transfers and extreme particle acceleration take place \citep{dermer10}. Thanks to the Large Area Telescope (LAT) aboard the \textit{Fermi Gamma-ray Space Telescope} (\textit{Fermi}; \citealt{atwood09}) which has been conducting a sky survey at energies above 100 MeV, we have an unprecedented map of the $\gamma$-ray sky \citep{0fgl,1fgl}. In its first 11 months of operation \textit{Fermi} detected and characterized 1451 $\gamma$-ray objects in the sky, culminating in the \textit{Fermi} LAT First Source List (1FGL; \citealt{1fgl}) and the First LAT AGN Catalog (1LAC; \citealt{1lac}). These detailed maps of the $\gamma$-ray sky are a rich dataset which that be used in cross-correlation analyses to shed light on the origin of energetic cosmic rays.

\citet{mirabal10} recently investigated the correlation of the complete sample of 1FGL sources with the PAO dataset (as of \citealt{palong}). Their analysis was carried out without any redshift cut-off or type selection on the 1FGL, and counted associations when PAO events occur within circles of radius $3.1^\circ$ around the \textit{Fermi} sources. \citet{mirabal10} found no evidence that UHECRs are associated with \textit{Fermi} sources for the particular angular radius considered. This does not imply though that there is no association at all between $\gamma$-ray sources and UHECRs. Considering that the 1FGL catalog is a ``mixed bag'' of $\gamma$-ray sources with quite different properties, if one particular class of $\gamma$-ray emitters is responsible for producing UHECRs, then other sources in the catalog bearing no relation with the cosmic rays might introduce ``noise'' in the correlation analysis. Furthermore, the correlation can be weak at a given angular separation and stronger at others.

In this work we use the 1FGL and 1LAC catalogs produced by the \textit{Fermi} LAT to investigate the correlation between the positions of $\gamma$-ray sources and the arrival directions of UHECRs measured by the PAO. Our goal is to unveil if specific types of $\gamma$-ray sources might be driving any correlation and at what angular separations these correlations become significant. In \textsection \ref{sec:data} we describe the datasets we use. In \textsection \ref{sec:method} we describe our cross-correlation method. In \textsection \ref{sec:results} we list the results of our cross-correlation analysis when applied to the different subsamples of the 1FGL and 1LAC catalogs. We discuss the implications of our results in \textsection \ref{sec:disc}. In particular, we list the potential ``$\gamma$-ray loud'' UHECR accelerators and discuss their properties in \textsection \ref{sec:sites}. Section \ref{sec:end} closes with our concluding remarks.

\section{Data}	\label{sec:data}

We base our analysis on the distribution of arrival directions of UHECRs with energies exceeding $5.7 \times 10^ {19} \ {\rm eV}$, collected  by the surface array of PAO between 1 January, 2004 and 31 August 2007, with an integrated exposure of $9.0 \times 10^3 \ {\rm km}^2 \ {\rm sr \ year}$ \citep{pa, palong}. This data set corresponds to 27 events with zenith angles smaller than $60^\circ$ and angular resolution of $\approx 1^\circ$ \citep{padirs}.

To search for the potential astrophysical sources of UHECRs, we use the \textit{Fermi} LAT First Source List (1FGL; \citealt{1fgl}) and the First LAT AGN Catalog (1LAC; \citealt{1lac}) of $\gamma$-ray sources produced after the first eleven months of operation of \textit{Fermi}. Source detection is based on the average flux over the 11-month period corresponding to a statistical significance higher than $4\sigma$. The 1FGL consists of the 1451 sources detected and characterized in the 100 MeV to 100 GeV range. A subset of the 1FGL consisting of 671 sources constitutes the 1LAC. 

1037 sources in the 1FGL are within the PAO field of view. \citet{1fgl} provides identifications or plausible associations of these $\gamma$-ray sources with objects in other astronomical catalogs: 93 of them correspond to Galactic sources (pulsars, pulsar wind nebulae, supernova remnants, X-ray binaries and globular clusters) and 453 to AGNs in the 1LAC. Of the 1LAC AGNs, 367 are blazars, 69 are AGNs of uncertain type and 17 are non-blazar AGNs that include, for example, the radio galaxies M87, Centaurus A and 3C 207.0 \citep{1lac}. 490 1FGL sources could not be associated with any counterpart. Most of the unassociated sources are near the Galactic plane, and could not be identified because of a combination of Galactic extinction and source confusion \citep{0fgl, 1fgl, 1lac}.

\begin{figure}
\centering
\includegraphics[scale=0.75]{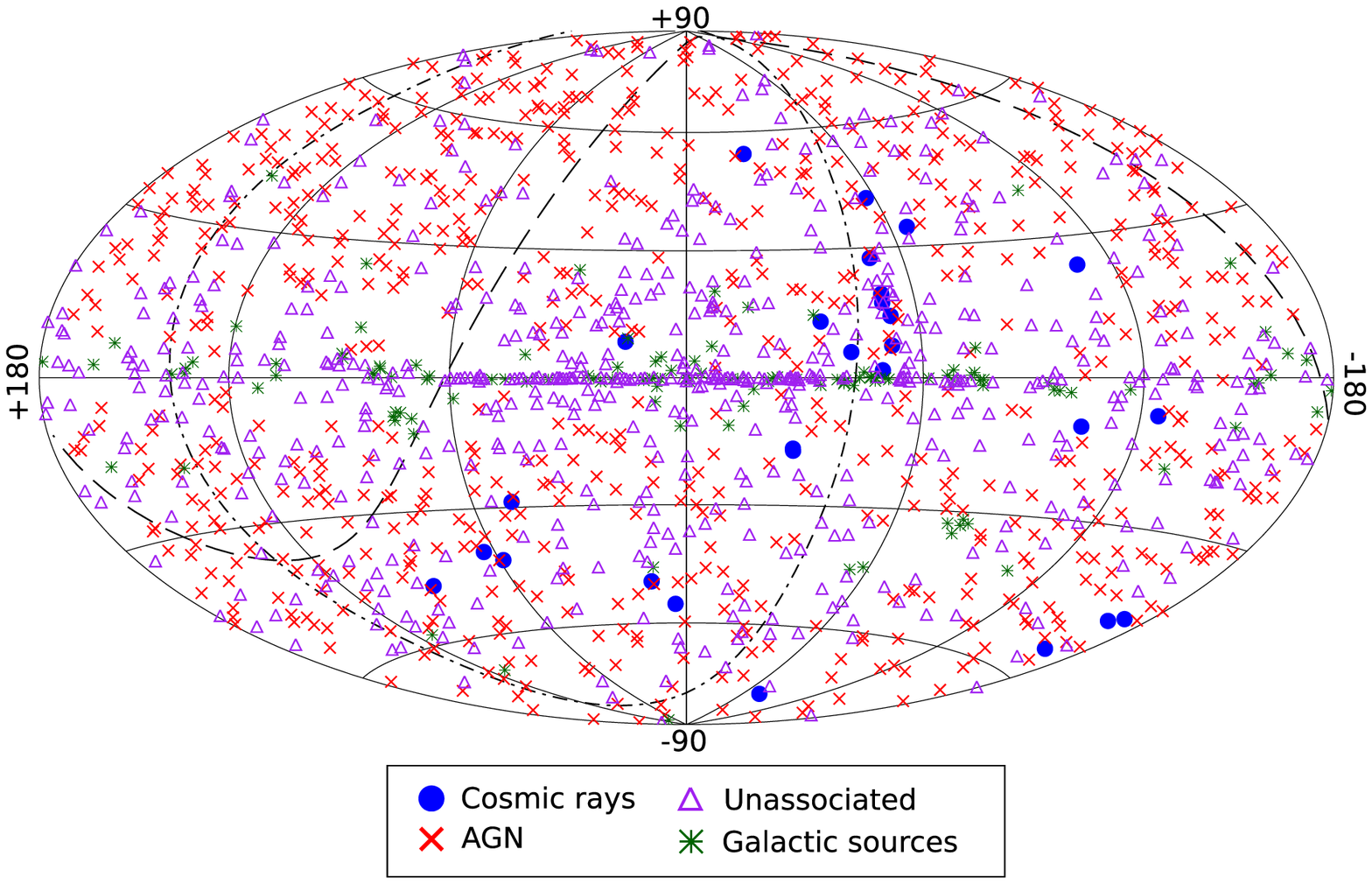}
\caption{Aitoff projection of the celestial sphere in Galactic coordinates, showing the 27 highest energy cosmic rays detected by the \textit{Pierre Auger Observatory} and the 1451 1FGL $\gamma$-ray sources observed with \textit{Fermi} (1037 in the field of view of PAO). The dashed line represents the border of the of the field of view of PAO, corresponding to a zenith angle of $60^\circ$. The dot-dashed line corresponds to the supergalactic plane. [See the electronic edition of the Journal for a color version of this figure.]}
\label{aitoff}
\end{figure}

Figure \ref{aitoff} shows the sky map in Galactic coordinates of the UHECRs events together with the 1FGL and 1LAC sources. Also shown is the supergalactic plane along which nearby galaxies cluster \citep{lahav00}.

\section{Cross-correlation method}	\label{sec:method}

In order to quantify the statistical cross-correlation between the \textit{Fermi} and PAO data, we compute the cumulative number of associated cosmic ray events as a function of angular separation $\psi$ (hereafter CCF for cumulative correlation function) from the data sets. The CCFs are computed by drawing circles of increasing radius around the \textit{Fermi} sources, and counting the number of cosmic ray events that occur within such circles. This is qualitatively similar to the approach of \citet{stephen05} for the correlation analysis between two different X-ray catalogs. 

The significance of the measured CCF (number of correlations measured in the observed data $N_o$ as a function $\psi$) is evaluated by comparing it against that expected by chance correlations assuming an isotropic flux of UHECRs. The chance CCF -- number of correlations expected by chance $N_c$ vs. $\psi$ --  is obtained by building simulated sets of UHECRs with the same number of observed events, drawn from an isotropic distribution of arrival directions convolved with the exposure function of the PAO \citep{sommers01}. 

For a given configuration of \textit{Fermi} sources and a given $\psi$, our null hypothesis is that the number of associations measured in the data is the same as the average number expected by chance: $N_o=\langle N_c \rangle$. We calculate the p-value $P$ as the probability that $N_c$ exceeds $N_o$. In practice, this p-value can be interpreted as the likelihood that the associations measured in the data could have occurred by chance if the null hypothesis is true, i.e. the likelihood that the observed configuration of $\gamma$-ray sources and UHECRs is a random accident. Assuming that the distribution of $N_c$ values obtained from the Monte Carlo simulation is normal, we define $P$ as
\begin{equation}
P(\psi) = \left\{ 
\begin{array}{l l}
  2 \left[ 1-\Phi ( \Delta N / \sigma ) \right], & \quad \Delta N \geq 0 \\
  2 \Phi ( \Delta N / \sigma ), & \quad \Delta N < 0 \\ 
\end{array} \right.
\end{equation}
where $\Delta N (\psi) \equiv N_o- \langle N_c \rangle$, $\sigma (\psi)$ is the standard deviation of the distribution of values of $N_c$ and $\Phi(x)$ is the cumulative distribution function for a normal distribution.

When describing our results in the sections that follow, we will often refer to the statistical significance level with which we can reject the null hypothesis. We will refer to this significance level as $S=\sqrt{2} \ {\rm erf}^{-1} (1-P)$, given in standard deviations. $S$ is just another way of expressing the p-value $P$.

\subsection{Consistency check}

\begin{figure}
\centering
\includegraphics[scale=1]{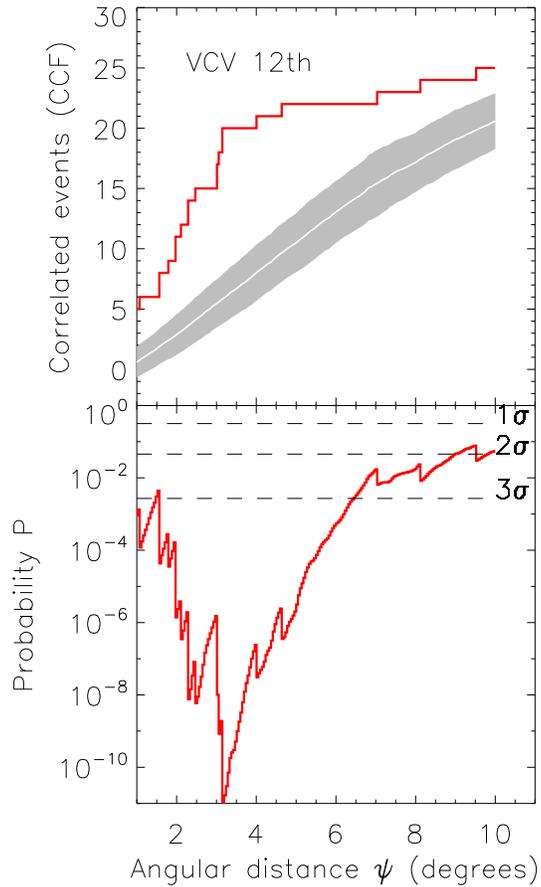}
\caption{Correlation signal between the arrival directions of UHECRs and the AGNs with $z \leq 0.017$ within the field of view of PAO in the 12th edition of the V\'eron-Cetty and V\'eron catalog. \textit{Upper panel:} cumulative number of cosmic ray events as a function of angular distance $\psi$ (CCF; black line). The average chance expectation with the 68\% confidence intervals is represented by the shaded region. \textit{Lower panel:} the solid line correspond to the probability $P$ that the number of observed correlated events is due to chance correlations. The dashed lines indicate the probabilities corresponding to the different significance levels for the null hypothesis.}
\label{fig:vcv12}
\end{figure}

As a consistency check of the method outlined above, we applied it to calculate the cross-correlation between the arrival directions of UHECRs and the 292 AGNs with redshift $z\leq0.017$ within the PAO field of view listed in the 12th edition of the V\'eron-Cetty and V\'eron (VCV) catalog \citep{vc}, previously studied by \citet{pa,palong}. 
Figure \ref{fig:vcv12} shows the resulting correlation signal. The upper panel in Figure \ref{fig:vcv12} shows the measured CCFs, with angular separations ranging from $1^\circ$ (approximately the $1\sigma$ uncertainty in the arrival directions of PAO cosmic rays) up to $10^\circ$. The lower panels display the probability $P$. 

Figure \ref{fig:vcv12} illustrates that $P$ reaches a minimal value of $8.3 \times 10^{-12}$  at the angular separation $\approx 3.2^\circ$, with 20 cosmic ray events correlated with AGNs at this angular scale, while only 5.9 correlations are expected by chance. Therefore, the results of our consistency check confirm with at least a significance level $S=6.5\sigma$ the anisotropy of the arrival directions of UHECRs and are in excellent agreement with those obtained by the statistical method used by the Pierre Auger Collaboration \citep{pa,palong}. This lends strong support to the validity of our statistical method to probe the degree of correlation between the PAO and \textit{Fermi} data sets.

As a matter of fact, when we apply our statistical test to the revised AGN sample in the 13th edition of the VCV catalog using the same redshift cutoff and considering only AGNs within the PAO field of view, we find that the association with UHECRs is maximized at $\psi \approx 2.3^\circ$ with $P_{\rm min} = 6.7 \times 10^{-10}$ and correspondingly $S_{\rm max} = 6.2\sigma$. Therefore, the correlation signal slightly changes when cross-correlating PAO events with the updated VCV catalog.

\section{Results}	\label{sec:results}

We applied the above method to obtain the cross-correlation signal between the 1FGL/1LAC and PAO sources. In the following subsections we analyse the correlation signal of PAO events with: (1) all sources in the 1FGL catalog (\textsection \ref{sec:1fgl}), (2) only to 1FGL Galactic sources (\textsection \ref{sec:gal}) and (3) only to 1LAC AGNs (Sections \ref{sec:agn} and \ref{sec:gzk}). In \textsection \ref{sec:agn} we impose no restriction on the redshifts of the sources while in \textsection \ref{sec:gzk} we restrict ourselves only to nearby AGNs within the GZK horizon.

\subsection{UHECRs vs. all 1FGL sources} \label{sec:1fgl}

\begin{figure}
\centering
\includegraphics[scale=1]{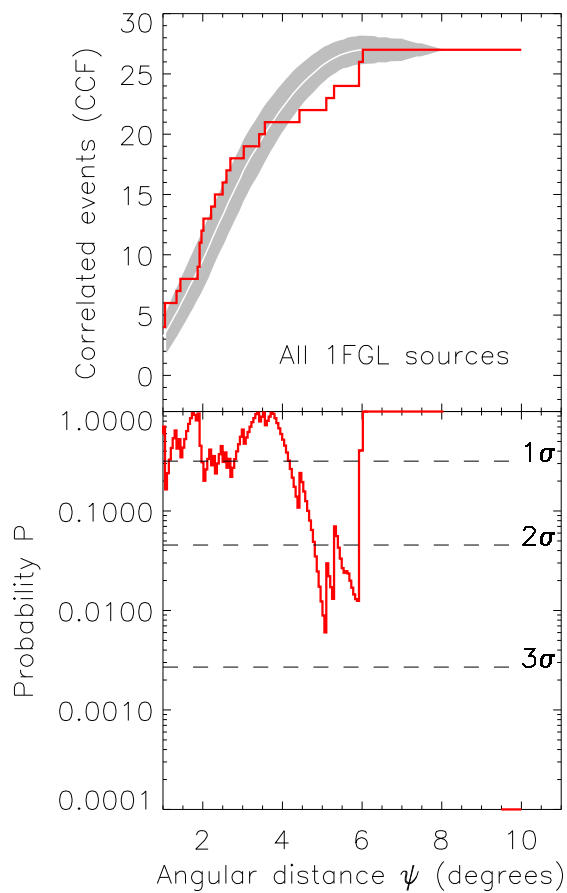}
\caption{Correlation signal between all 1FGL $\gamma$-ray sources and energetic cosmic ray events. \textit{Upper panel:} cumulative number of cosmic ray events as a function of angular distance $\psi$ (CCF; black line). The average chance expectation with the 68\% confidence intervals is represented by the shaded region. \textit{Lower panel:} the solid line correspond to the probability $P$ that the number of observed correlated events is due to chance correlations. The dashed lines indicate the probabilities corresponding to the different significance levels for the null hypothesis.}
\label{fig:all}
\end{figure}

Figure \ref{fig:all} shows the correlation signal resulting from comparing the positions of all 1037 1FGL $\gamma$-ray sources within the PAO field of view with the directions of energetic cosmic ray events. The upper panel shows the measured CCFs, with angular separations ranging from $1^\circ$ (approximately the $1\sigma$ uncertainty in the arrival directions of PAO cosmic rays) up to $10^\circ$. The lower panels display the probability $P$. 

Both panels of Figure \ref{fig:all} show that the number of observed correlated events follows the pattern of isotropic expectations within the $1\sigma$ significance until $\psi \approx 4^\circ$. In the range of angular separations $4^\circ \lesssim \psi < 6^\circ$, there is a systematic paucity of measured correlations with respect to chance expectations with $S>1\sigma$. In particular, the maximal ``anti-association'' between the 1FGL and PAO datasets corresponds to the global minimum of the probability signal $P_{\rm min}=0.6 \%$, which occurs at $\psi \approx 5.1^\circ$. This minimal probability corresponds to the significance level $S=2.7\sigma$. At $\psi \approx 5^\circ$, 22 PAO events correlate with all 1FGL sources while 25.8 are expected from an isotropic flux.

The case corresponding to circles of radius $3.1^\circ$ deserves special attention. This is the estimated angular radius around AGNs in the V\'eron-Cetty and V\'eron catalog that maximizes their correlation with the arrival directions of UHECRs \citep{pa}. Figure \ref{fig:all} illustrates that at $\psi = 3.1^\circ$ we find a high probability of chance correlation, $P=0.6$ ($S=0.5\sigma$). This agrees with the analysis of \citet{mirabal10}.

The 1FGL catalog consists of a ``mixed bag'' of Galactic and extragalactic $\gamma$-ray sources with quite different properties. In the subsections below, we investigate the correlation of the individual classes of $\gamma$-ray sources identified by \citet{1fgl} with the positions of PAO events.

\subsection{UHECRs vs. 1FGL Galactic sources} \label{sec:gal}

\begin{figure}
\centering
\includegraphics[scale=1]{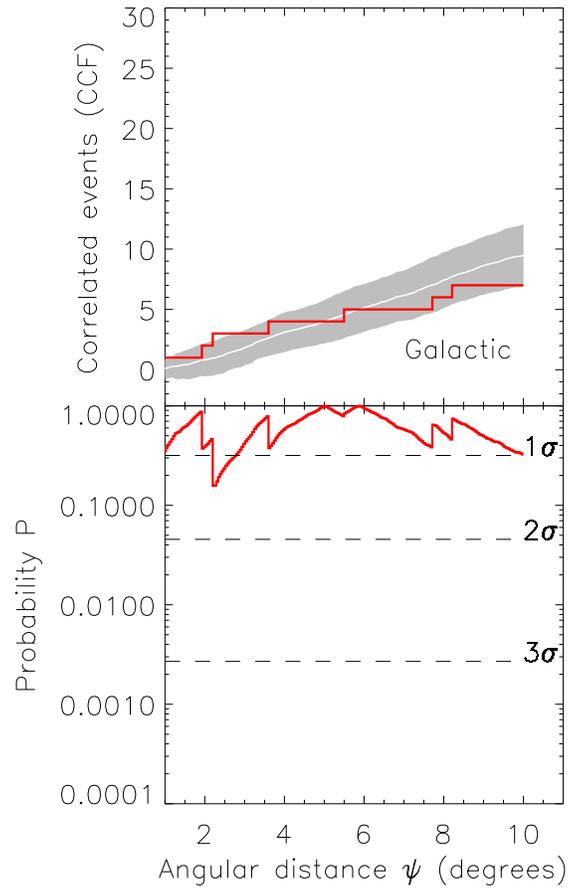}
\caption{Same as Figure \ref{fig:all} for the 1FGL Galactic sources.}
\label{fig:gal}
\end{figure}

Figure \ref{fig:gal} shows the correlation signal measured in the subset of 93 1FGL Galactic sources. The number of observed correlated events follows well the pattern of an isotropic expectation within the $1\sigma$ significance. There is a narrow fluctuation in the signal above $S=1\sigma$ at $\psi \approx 2.4^\circ$. Nevertheless, there are only 3 correlated events at this angular distance with 1 correlation expected from isotropy.

\subsection{UHECRs vs. 1LAC AGNs} \label{sec:agn}

In this section we study the likelihood of association between UHECRs and the 453 AGNs identified in the 1FGL, i.e. the sources in the 1LAC catalog within the PAO field of view. Figure \ref{fig:1lacmap} shows the sky map in Galactic coordinates of the PAO events and the 1LAC sources. These AGNs are located at distances ranging from $z \approx 0.001$ (the nearest 1LAC AGN is NGC 253) up to $z \approx 3.2$, which corresponds to the FSRQ blazar PKS 0336-017 \citep{1lac}.

\begin{figure}
\centering
\includegraphics[scale=0.7]{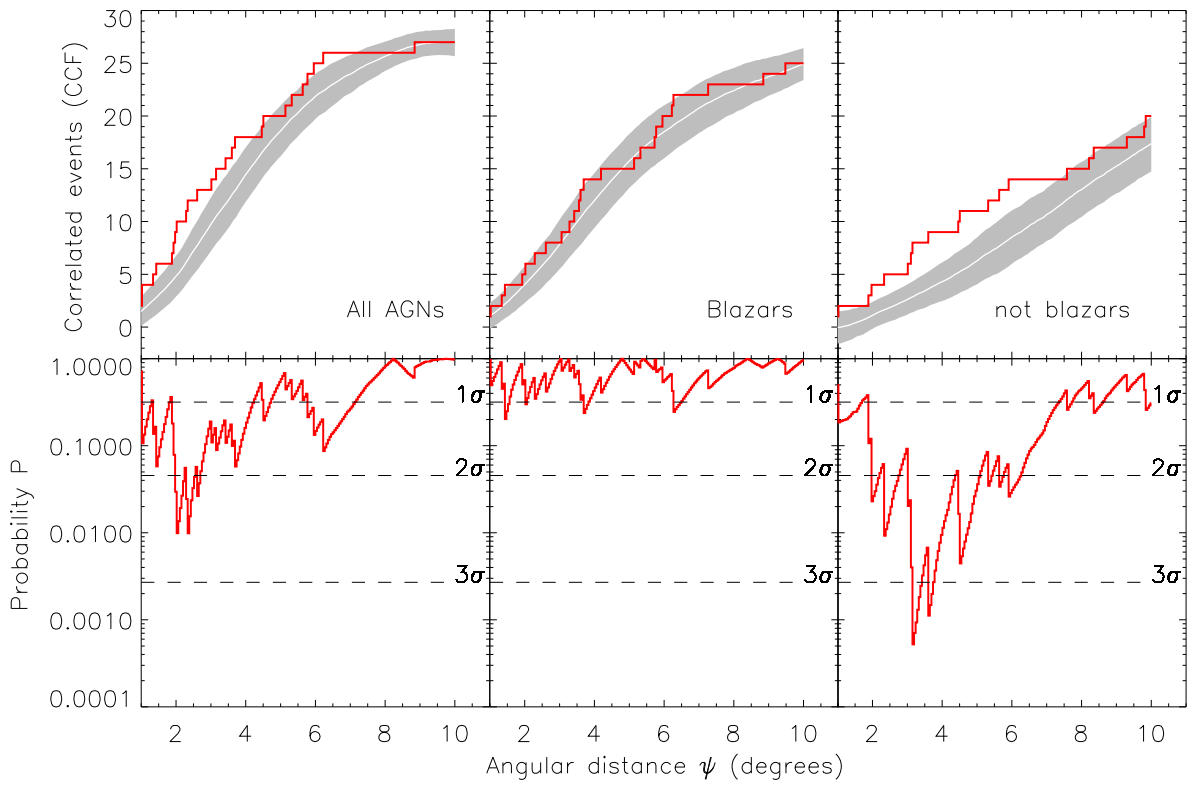}
\caption{Same as Figure \ref{fig:all} for the 453 1LAC AGNs.}
\label{fig:agn}
\end{figure}

The left panel of Figure \ref{fig:agn} shows the correlation signal of the PAO events with all 1LAC AGNs. There is a maximal association between the two data sets at the angular scale $\psi \approx 2.4^\circ$ (i.e. $P$ reaches a minimal value $P_{\rm min} \approx  0.01$, corresponding to $S_{\rm max}=2.6\sigma$). Within $2.4^\circ$, 12 events correlate with all the \textit{Fermi} AGNs without distinction of distance or AGN class, whereas 6.5 correlations are expected by chance. 

The middle panel of Figure \ref{fig:agn} shows the correlation signal with the 367 1LAC blazars. There is no significant correlation (i.e. $S<1\sigma$) between blazars and UHECRs that cannot be explained by an isotropic flux of cosmic rays.
We also study the amount of correlation obtained when cross-correlating separately the two classes of blazars, BL Lacertae and FSRQ, with the PAO UHECRs. We obtained for the 169 BL Lacertae 1LAC sources a quite similar correlation signal to that of all blazars (i.e. consistent with chance correlations). If we select only FSRQ sources, we have a sample of 198 FSRQ objects, which increases the significance of the correlation up to $2.3\sigma$ at $\psi \approx 6.5^\circ$. 
It is interesting to note that 1LAC FSRQ blazars are located at $z>0.15$, well outside the GZK horizon. Therefore,  we should not expect an association of these objects with the arrival directions of UHECRs.

The right panel of Figure \ref{fig:agn} displays the amount of cross-correlation between the subset of 86 AGNs consisting of non blazars (i.e. 17 misaligned jet sources such as the radio galaxies Centaurus A and M87) together with the AGNs that have not been identified yet (69 sources). The lower panel shows that the correlation significance raises above $1\sigma$ in the range $2^\circ \lesssim \psi \lesssim 7.5^\circ$.  In particular, the probability signal reaches the minimum value $P_{\rm min} \approx 5 \times 10^{-4}$ ($S_{\rm max}=3.5\sigma$) at $\psi \approx 3.2^\circ$, with 8 observed PAO event correlations, whereas only 2.9 chance correlations are expected. These PAO events are correlated with 7 AGNs, of which Cen A, 4C+04.77 and NGC 4945 are the only sources whose AGN types are known. 

\subsection{UHECRs vs. nearby 1LAC AGNs} \label{sec:gzk}

Protons or nuclei with energies above 60 EeV interact with the cosmic microwave background and suffer the GZK effect: a strong attenuation of their flux for distant sources \citep{g,zk}. This defines a GZK horizon located roughly at a distance $\sim 100-200$ Mpc \citep{harari06}. Astrophysical sources located outside the GZK horizon are not expected to contribute appreciably to the observed flux of energetic cosmic rays. Motivated by this expectation, we investigate the correlation of PAO events with 1LAC AGNs located within the GZK horizon. 
We define the GZK horizon as being located at a comoving distance of 200 Mpc, which corresponds to $z_{\rm max}=0.048$, assuming the $\Lambda$CDM cosmology with $H_0  = 71 \ \textrm{km} \ \textrm{s}^{-1} \ \textrm{Mpc}^{-1}$, $\Omega_m  = 0.270$, $\Omega_{\rm \Lambda}  = 0.730$ and $\Omega_k=0$ \citep{wmap}.

\begin{figure}
\centering
\includegraphics[scale=0.75]{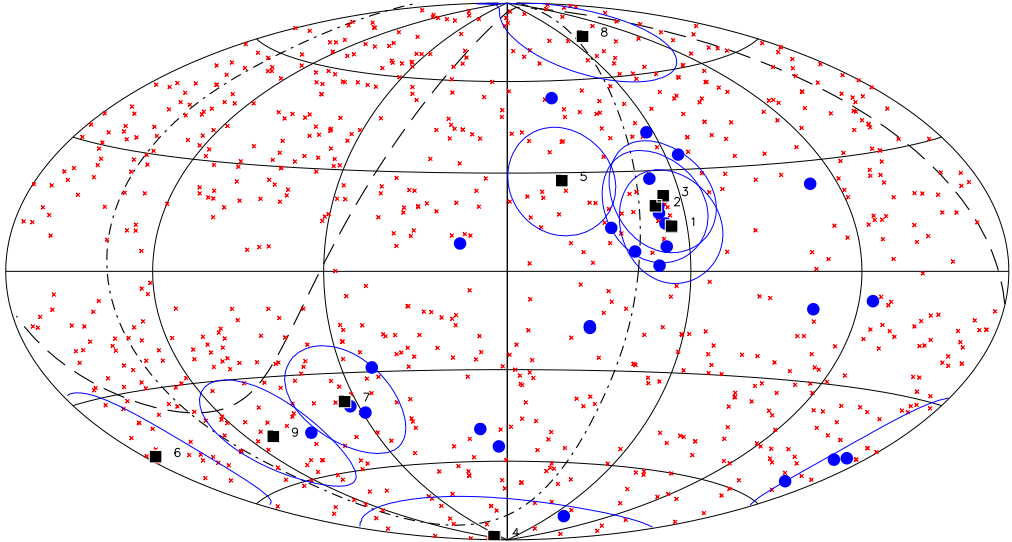}
\caption{Aitoff projection of the celestial sphere in Galactic coordinates, showing the PAO events (blue circles) and the 1LAC AGNs. The square symbols represent the AGNs with $z \leq 0.048$ (i.e. within the GZK horizon) in the PAO field of view with circles of radius $16.9^\circ$ centered at each square, while the x symbols denote all other AGNs. The numbers can be used to identify the 1LAC GZK AGNs in Table \ref{tab:agn}. The dashed and dot-dashed lines follow the notation of Figure \ref{aitoff}. [See the electronic edition of the Journal for a color version of this figure.]}
\label{fig:1lacmap}
\end{figure}

We use the redshift information provided in the 1LAC \citep{1lac} to discard the AGNs within the PAO field of view for which $z>z_{\rm max}$. Only 9 AGNs remain, including the radio galaxies Centaurus A (the nearest radio galaxy) and M87, the Seyfert 2 NGC 4945, the Seyfert 1 ESO 323-G77 and the starburst galaxy NGC 253 (as previously noted by \citealt{dermer10}). Only three AGNs within the GZK horizon are BL Lacertae objects. There is no FSRQ AGN within the GZK horizon, with the nearest FSRQ blazar in the 1LAC being located at $z=0.15$. 1LAC sources inside the GZK horizon are shown as squares in Figure \ref{fig:1lacmap}.

Figure \ref{fig:gzk} shows the correlation signal between 1LAC AGNs with $z \leq 0.048$ and UHECRs, with $\psi$ reaching up to $20^\circ$. The noteworthy feature here is that there are three characteristic angular scales where the probability of the observed configuration being explained by an isotropic flux of cosmic rays is minimized: $\psi_1 \approx 2.3^\circ$, $\psi_2 \approx 6.5^\circ$ and $\psi_3 \approx 16.9^\circ$. 
The first local minimum of the probability signal located at $\psi_1 \approx 2.3^\circ$ corresponds to only 4 associations, with $P(\psi_1)=6 \times 10^{-5}$ and $S(\psi_1)=4\sigma$. 
The second local minimum of the probability signal located at $\psi_2 \approx 6.5^\circ$ corresponds to 7 measured correlations whereas 0.9 chance correlations are expected, with $P(\psi_2)=5 \times 10^{-7}$ and $S(\psi_2)=5\sigma$. 
The absolute minimum of $P$ is characterized by $\psi_3 \approx 16.9^\circ$, $N_o=17$, $N_c=5.7$, $P(\psi_3)=5 \times 10^{-8}$ and $S(\psi_3)=5.4\sigma$.

Within the distance $\psi_1$, 4 PAO events correlate with 3 AGNs: NGC 4945, Centaurus A and 4C+04.77; at $\psi_2$, 7 PAO events correlate with 4 AGNs: NGC 4945, Centaurus A, ESO 323-G77 and 4C+04.77; at $\psi_3$, 17 events are associated with 7 AGNs: NGC 4945, Centaurus A, ESO 323-G77, 4C+04.77, NGC 253, NGC 1218 and RXJ0008.0+1450. Table \ref{tab:agn} lists the 9 1LAC AGNs located inside the GZK horizon, including their 1FGL names, Galactic coordinates (longitude $l$ and latitude $b$), identifications and redshifts according to \citet{1fgl,1lac}. We also list the number of PAO events potentially correlated with each AGN at the angular scales $\psi_2$ and $\psi_3$. The ellipses in Figure \ref{fig:1lacmap} correspond to circles of radius $\psi_3$ centered on the GZK AGNs.

\section{Discussion}	\label{sec:disc}

We begin by pointing out some caveats in our analysis of the correlation of 1LAC AGNs and PAO events. Firstly, the 1LAC catalog of AGNs is increasingly incomplete towards the Galactic plane (\citealt{1fgl,1lac}; see Figure \ref{fig:1lacmap}) due to a combination of the effects of dust extinction in the Milky Way and source confusion. These effects may have some impact on our estimate of the correlation strength, since we might be missing some AGNs lying close to the Galactic plane.
Secondly, there are many AGNs in the 1LAC that have not yet been identified with counterparts in other wavelengths, and lack redshift determinations \citep{1lac}. This may affect our cross-correlation analysis using the subset of 1LAC GZK AGNs, since we might be missing AGNs inside the GZK horizon that are yet unidentified in the 1LAC.
With the progressive increase in the exposure time of the \textit{Fermi} LAT survey observations, future releases of the \textit{Fermi} LAT source catalog will improve on both these issues: they will increase the number of AGN detections near the Galactic plane and increase the number of associations of $\gamma$-ray sources with multiwavelength counterparts \citep{0fgl,1fgl,1lac}.

It is interesting to point out that $\psi=3.2^\circ$ -- the angular separation which maximizes the association of PAO events with the AGNs in the 12th VCV catalog -- does not correspond to a local minimum of the probability signal in our analysis of the correlation of 1LAC GZK sources and PAO events, even though at $3.2^\circ$ the hypothesis of the observed correlations being due to chance can be rejected at the $4.2\sigma$ level. On the other hand, the radius $2.3^\circ$ that maximizes the association with the sources in the 13th VCV catalog corresponds to a local minimum of the correlation signal in Figure \ref{fig:gzk}.

We find no correlation between 1LAC blazars and UHECRs when considering all blazars detected by Fermi/LAT up to $z \approx 3.2$. In other words, relativistic jets nearly pointed along our line of sight do not seem to be significantly associated with the arrival directions of UHECRs on any angular scale. This result is in disagreement with the controversial claims of correlation of HiRes data with BL Lac objects \citep{gorbunov04}, but is expected on theoretical grounds \citep{vp09}.

\subsection{The potential UHECR accelerators}	\label{sec:sites}

Our results pinpoint 7 ``$\gamma$-ray loud'' 1LAC AGNs as the potential sites for the production of PAO UHECRs (see Table \ref{tab:agn}): Cen A, NGC 4945, ESO 323-G77, 4C+04.77, NGC 1218, RX J0008.0+1450 and NGC 253 (ordered by increasing number of associated UHECRs at the angular distance $16.9^\circ$). None of these AGNs display significant variability in their $\gamma$-ray emission \citep{1fgl,1lac}. The three sources Cen A, NGC 4945 and ESO 323-G77 dominate the cosmic ray budget at all angular distances, having many UHECRs clustered near them. We discuss the likelihood of production of UHECRs in each of these sources.

\begin{figure}
\centering
\includegraphics[scale=1]{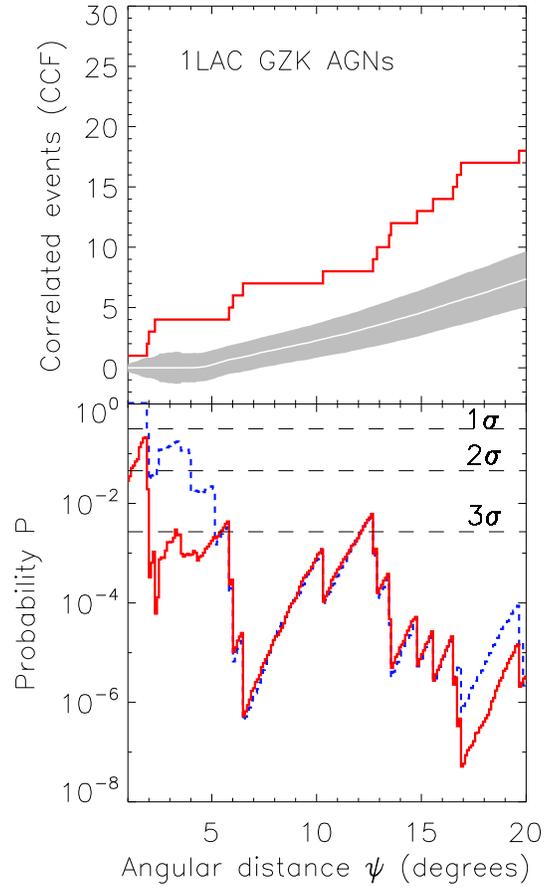}
\caption{Same as Figure \ref{fig:all} for the 9 1LAC AGNs within the GZK horizon. In the lower panel, the solid and dashed lines correspond to the correlation signal including and not including Cen A, respectively.}
\label{fig:gzk}
\end{figure}

As pointed out before by different authors, the nearby radio galaxy Centaurus A is an excellent candidate for the acceleration of UHECRs in its jets and radio lobes \citep{moska09, hardcastle09, honda09, rieger09, dermer10, piran10}. Cen A is a prominent $\gamma$-ray source \citep{1fgl,1lac,dermer10}, being associated with two UHECRs \citep{palong} and potentially even more events \citep{gorbunov08, nagar08, moska09}. We can test the impact of Centaurus A in our results by removing it from the 1LAC data and recalculating the correlation signal of nearby 1LAC AGNs without Cen A. The dashed line in the lower panel of Fig. \ref{fig:gzk} corresponds to the correlation signal without Cen A, which differs from the signal including Cen A (solid line in Fig. \ref{fig:gzk}) in the following way: the strength of the correlation is considerably weakened on separations smaller than $\approx 5^\circ$, the minimum of the probability signal at $\psi_2$ essentially does not change its associated probability and the minimum at $\psi_3$ has its significance slightly decreased to $5\sigma$. Therefore, statistically speaking, the impact of Cen A on the correlation is mainly in the significance of the short-distance associations. 

NGC 4945 is classified as either a starburst galaxy or an obscured Seyfert 2 nucleus showing no signs of the presence of relativistic outflows (e.g., \citealt{spoon00, chou07}), while ESO 323-G77 is a Seyfert 1 galaxy that displays evidence for the presence of a mildly relativistic outflow in its X-ray spectrum \citep{jimenez08}. We have to be careful when associating these galaxies with the nearby PAO events, since neither of them display powerful outflows or are energetically impressive (but see \citealt{boldt99,peer09}). Both galaxies are located quite close to Cen A and any events associated with Cen A could also be mistakenly attributed to NGC 4945 and ESO 323-G77 in a correlation analysis, given their proximity (i.e. a case of accidental correlation). 

A better understanding of the nature of the $\gamma$-ray emission in NGC 4945 and ESO 323-G77 could shed light on the issue of their likelihood as accelerators of energetic cosmic rays. Given the available information on these AGNs, our analysis suggests that Centaurus A is the primary candidate for the production site of several UHECRs observed in the direction of its group of galaxies. We note that ESO 323-G77 shows evidence of the presence of a relativistically broadened Fe K$\alpha$ \citep{jimenez08}, which suggests that the supermassive black hole in this galaxy could be rapidly spinning, with the Kerr black hole being a potential UHECR accelerator \citep{boldt99, vp08, dermerbook}.

\subsection{Angular separations vs. deflection angles}	\label{sec:angles}

The correlation signal obtained in this work will be helpful to constrain the results of simulations of the propagation of energetic cosmic rays originating from extragalactic sources \citep{palong, lima09, ryu10}. One should be careful when mapping the separation angle $\psi$ that minimizes $P$ for a set of reference sources into the actual deflection angles $\theta$ suffered by the cosmic rays on their way towards Earth with respect to their actual sources. For instance, if the 1LAC GZK sources that we studied are tracing the production sites of the observed PAO events, which is a tempting possibility given the high significances of association that we obtain, the double-peaked significance pattern in the correlation between 1LAC GZK sources and PAO events raises a challenging question: Which of the angular scales $\psi_1 \sim 2^\circ$, $\psi_2 \sim 7^\circ$ or $\psi_3 \sim 17^\circ$ actually corresponds to a deflection angle? The answer to this question is beyond the scope of this work, but it is worthwhile to compare our results with numerical simulations.

\citet{ryu10} modeled the propagation of UHECRs from source AGNs to observers through the intergalactic magnetic field, using a model universe based on results from cosmological structure formation simulations.  Interestingly enough, they obtained the mean value of deflection angle between the arrival directions of UHECRs and the actual sources of $\langle \theta \rangle \sim 14^\circ - 17.5^\circ$. Therefore, our inferred angular distance $\psi_3 \sim 17^\circ$ -- which corresponds to the maximum significance $S(\psi_3) = 5.4\sigma$ (see Figure \ref{fig:gzk}) -- is within the theoretical range of mean deflection angles calculated by \citet{ryu10}!

On the other hand, the separation angle between the arrival directions of UHECRs and the position of the nearest ``mock'' AGN in the simulation of \citet{ryu10} is $\langle \psi_{\rm sim} \rangle \sim 3.5^\circ - 4^\circ$, therefore $\langle \theta \rangle \gg \langle \psi_{\rm sim} \rangle$. Quite interestingly, when comparing the position of PAO events against those of all 1LAC AGNs (left panel of Fig. \ref{fig:agn}) we find a much smaller separation angle that minimizes the probability of the observed configuration being due to an isotropic UHECR flux, $\psi \approx 2.4^\circ$. Taken together, our results seem to be in rough agreement with the theoretical results of \citet{ryu10}.

\subsection{Relation to the supergalactic plane}

The SGP traces the distribution of matter in the local universe, with several nearby galaxy clusters and AGNs being concentrated near this plane \citep{lahav00}, as well as the arrival directions of many UHECRs \citep{pa, palong}. For instance, 13 out of the 27 (48\%) PAO events are located within a distance of $15^\circ$ to the SGP, with the average distance of UHECRs to the SGP of $\approx 24.5^\circ$. 

Of the 9 1LAC AGNs with $z \leq 0.048$ in the PAO field of view, the 7 objects that correlate with energetic PAO cosmic rays within $\psi_3=16.9^\circ$ are located at an average distance to the SGP of $\langle \psi_{\rm sgp}^{\rm agn} \rangle = 16.8^\circ$. For comparison, PAO events associated with the 1LAC GZK AGNs are located at the average distance $\langle \psi_{\rm sgp}^{\rm PAO} \rangle = 19.7^\circ$. This suggests that the nearby $\gamma$-ray bright AGNs associated with UHECRs on average fall very close to the nearby local overdensity of matter in the universe, even though the general distribution of 1LAC AGNs does not follow noticeably the SGP. 

\section{Concluding remarks}	\label{sec:end}

We analysed the correlation between the positions of the $\gamma$-ray sources in the \textit{Fermi} LAT First Source Catalog (1FGL) and the First LAT AGN Catalog (1LAC) produced during the first eleven months of operation of the \textit{Fermi Gamma-ray Space Telescope}, and the arrival directions of the highest energy cosmic rays measured during 1.2 years of operation of the \textit{Pierre Auger Observatory}. Using our cross-correlation test, we are particularly interested in the value of separation angle $\psi$ between the positions of PAO events and reference sources that maximizes the significance level $S$ of rejecting the null hypothesis (that the observed configuration is due to chance assuming an isotropic flux of UHECRs). 
Our main results from the cross-correlation analysis of each 1FGL subset  (in the PAO field of view) with the PAO events are summarized below: 
\begin{itemize}
\item When considering all $\gamma$-ray sources in the 1FGL, $S_{\rm max}=2.7\sigma$ at $\psi \approx 5.1^\circ$ with less observed correlations than expected assuming the null hypothesis.
\item Galactic 1FGL sources are not significantly correlated with the arrival directions of energetic cosmic rays ($S<1\sigma$).
\item Considering the 1LAC AGNs we find $S_{\rm max}=2.6\sigma$ at $\psi \approx 2.4^\circ$.
\item We find no significant correlation with 1LAC blazars ($S<1\sigma$).
\item Considering the subset consisting of unidentified AGNs and non-blazars in the 1LAC, $S_{\rm max}=3.5\sigma$ at $\psi \approx 3.2^\circ$.
\item We find a significant association with 1LAC sources closer than 200 Mpc, with the significance peaking at the separations $\psi_1=2.3^\circ$ ($4\sigma$), $\psi_2=6.5^\circ$ ($5\sigma$) and $\psi_3=16.9^\circ$ ($5.4\sigma$).
\end{itemize}

For the subset of nine 1LAC AGNs closer than 200 Mpc ($z \leq 0.048$, roughly the distance to the GZK horizon), we reject with the highest significance the null hypothesis compared to the other subsets of 1FGL sources ($P_{\rm min}=5 \times 10^{-8}$ at $\psi_3 = 16.9^\circ$). 
Of the 27 PAO events, 17 are located within $16.9^\circ$ of 7 nearby 1LAC AGNs: Centaurus A, NGC 4945, ESO 323-G77, 4C+04.77, NGC 1218, RX J0008.0+1450 and NGC 253. These ``$\gamma$-ray loud'' AGNs are likely candidates for the production sites of UHECRs.

The considerable separation angle that maximizes $S$ could suggest that the trajectories of the particles on their way to Earth are significantly deflected by the intervening magnetic fields, consistent with the numerical simulations of \citet{ryu10}. This may set important constraints for the understanding of the structure of magnetic fields along the line of sight. 

Given the small number of currently detected UHECR events, this work represents a first step towards understanding the relation among some of the highest-energy phenomena in the universe detected by PAO and \textit{Fermi}.
The data set that the PAO and \textit{Fermi} will gather in the next few years should improve the correlation statistics of our work. Furthermore, cosmic ray observatories in the northern hemisphere such as the anticipated northern PAO site will offer a way of further identifying the sources of energetic cosmic rays. 
Thanks to PAO and \textit{Fermi}, new perspectives to understand the relation among the highest-energy phenomena in the universe are being made possible!

\acknowledgments

We are grateful to Graziela R. Keller, Tib\'erio Vale and Mike Brotherton for productive discussions and Brian Baughman for pointing out useful references. We also thank the anonymous referee for a number of suggestions that helped improve this work. We acknowledge support from the Brazilian institution CNPq, and thank the \textit{Fermi Science Support Center} for promptly making the 0FGL, 1FGL and 1LAC source lists available in electronic formats to the community. This research has made use of the NASA/IPAC Extragalactic Database (NED) which is operated by the Jet Propulsion Laboratory, California Institute of Technology, under contract with the National Aeronautics and Space Administration.

{\it Facilities:} \facility{Fermi}, \facility{PAO}.

\begin{deluxetable}{ccccccccc}
\tabletypesize{\scriptsize}
\tablecaption{1LAC AGNs with $z \leq 0.048$ and the associated PAO events
\label{tab:agn}}
\tablewidth{0pt}
\tablehead{
\colhead{1FGL} & \colhead{$l$} & \colhead{$b$} & \colhead{Class\tablenotemark{a}} & \colhead{Name\tablenotemark{b}} & \colhead{$z$} & \colhead{ID\tablenotemark{c}} & \colhead{$N(6.5^\circ)$\tablenotemark{d}} & \colhead{$N(16.9^\circ)$} \\
\colhead{name} & & & & & & & &
}
\startdata
J1305.4-4928  & $-54.7^\circ$ & $13.3^\circ$ & Sy 2 & NGC 4945 & 0.002 & 1 & 4 & 7 \\
J1325.6-4300  & $-50.4^\circ$ & $19.4^\circ$ & RG & Cen A & 0.002 & 2 & 3 & 7  \\
J1307.0-4030 & $-54^\circ$ & $22.4^\circ$ & Sy 1 & ESO 323-G77 & 0.015 & 3 & 3 & 6  \\
J0047.3-2512 & $97.4^\circ$ & $-88^\circ$ & SB & NGC 253 & 0.001 & 4 & 0 & 1 \\
J1517.8-2423 & $-19.3^\circ$ & $27.6^\circ$ & BLL & Ap Lib & 0.048 & 5 & 0 & 0  \\
J0308.3+0403 & $174.9^\circ$ & $-44.5^\circ$ & BLL & NGC 1218 & 0.029 & 6 & 0 & 2  \\
J2204.6+0442 & $64.7^\circ$ & $-38.6^\circ$ & Sy 1/BLL & 4C+04.77 & 0.027 & 7 & 2 & 3  \\
J1230.8+1223 & $-76.2^\circ$ & $74.5^\circ$ & RG & M87 & 0.004 & 8 & 0 & 0 \\
J0008.3+1452 & $107.6^\circ$ & $-46.7^\circ$ & Sy 1 & RX J0008.0+1450 & 0.045 & 9 & 0 & 1 \\
\enddata
\tablenotetext{a}{Meaning of acronyms: RG = radio galaxy, BLL = BL Lacertae, Sy 1 = Seyfert 1, Sy 2 = Seyfert 2, SB = starburst galaxy.}
\tablenotetext{b}{Name of identified or likely associated source.}
\tablenotetext{c}{Identification of each source numbered in Figure \ref{fig:1lacmap}.}
\tablenotetext{d}{Number of PAO events potentially associated with each \textit{Fermi} source at the given separation.}
\end{deluxetable}

\end{document}